\documentclass[
reprint,
bibnotes,
amsmath,amssymb,
aps,
prd
]{revtex4-2}

\usepackage{graphicx}
\usepackage{dcolumn}
\usepackage{enumitem}
\usepackage{bm}
\usepackage[many]{tcolorbox}
\usepackage{shuffle}
\usepackage{xcolor}
\usepackage{physics}
\usepackage{tabularx}
\usepackage{diagbox}
\usepackage{subcaption}
\usepackage{multirow}
\usepackage{caption}
\usepackage{tikz}
\usetikzlibrary{decorations.markings}
\usetikzlibrary{decorations.pathmorphing}
\usetikzlibrary{intersections}
\usetikzlibrary{calc}
\usepackage{verbatim}

\usepackage{CJK}
\definecolor{Maroon}{RGB}{128, 0, 0}
\definecolor{Green}{rgb}{0.0, 0.5, 0.0}
\definecolor{Purple}{rgb}{0.5, 0.0, 0.5}
\definecolor{Blue}{rgb}{0.0, 0.0, 0.6}

\begin{document}

\begin{CJK*}{UTF8}{}
\CJKfamily{gbsn}

\title{Correlators are simpler than wavefunctions}

\author{Nima Arkani-Hamed$^{1}$}
\email{arkani@ias.edu}
\author{Ross Glew$^{2}$}
\email{r.glew@herts.ac.uk}
\author{Francisco Vaz\~{a}o$^{1,}$}
\email{fvvazao@ias.edu}

\affiliation{
$^{1}$School of Natural Sciences, Institute for Advanced Study, Princeton, NJ 08540 \\
$^{2}$Department of Physics, Astronomy and Mathematics, University of Hertfordshire, Hatfield, Hertfordshire, AL10 9AB, UK
}

\begin{abstract}
Recent works reveal a simplicity in equal-time correlators absent from wavefunctions. We show that it follows from the fact that correlators are full spacetime integrals, rather than half-spacetime integrals as for the wavefunction. This makes striking properties manifest: some wavefunction singularities are absent, factorization simplifies, and around every pole the Laurent expansion has vanishing first subleading term. Near the total-energy pole, the expansion to second order is generated by a differential operator on scattering amplitudes.
\end{abstract}
\maketitle
\end{CJK*}

\section{Introduction}
Recently, the flat-space vacuum wavefunction for scalar theories with polynomial interactions has attracted considerable interest \cite{CosmoPolys,Cosmohedron,Anninos:2014lwa,CosmoBoot,DiffEq_CosmoCorr,KinFlow,GeometryCosmoCorr,Benincasa:2024lxe,Goodhew:2020hob,CosmoReview,KinFlowLoop,IRSub,MelvillePajer,PajerUnitarityLocality,GoodhewCutCosmoCorr,PokrakaDEs,CosmoSMatrix,Sachs2,CosmoTreeTheo,Meltzer,Bittermann,CespedesScott,Sachs3,Thavanesan:2025kyc,Goodhew:2024eup,Forcey:2025voc,Glew:2025ypb,HiddenZerosWF,De:2024zic,Glew:2025otn,Glew:2025ugf,Raman:2025tsg,Xianyu:2025lbk,Benincasa:2019vqr,Fan:2024iek,Figueiredo:2025daa}, due to its close connection with the cosmological wavefunction in simple toy model cosmologies. Defined by the path integral \cite{CosmoReview}:
\begin{align}
    \Psi [\Phi] &= \int_{\varphi(t=-\infty)=0}^{\varphi(t=0) = \Phi} \mathcal{D}\varphi\: e^{-i \int_{-\infty}^{0} dt\,\int_{-\infty}^{+\infty} d^n x\, S[\varphi]},
\end{align}
the wavefunction encodes the quantum state of the system on the $t=0$ slice. The physically relevant observables, however, are the equal-time correlation functions, computed by \cite{Donath:2024utn}:
\begin{align}
    \langle \Phi^N \rangle &= \int \mathcal{D}\Phi\:\prod_{i}^{N} \Phi(0,\vec{x}_i)\: |\Psi[\Phi]|^2.
\end{align}
Since the wavefunction serves only as an intermediate object for computing correlators, it is natural to expect that the correlators themselves may display simplifications that are otherwise absent in the wavefunction.

\begin{figure}[t]
\centering
\includegraphics[scale=0.75]{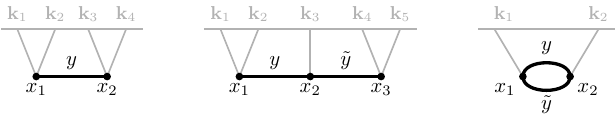}
\caption{Feynman graphs for $\Psi_G$.}
\label{figs:feyn}
\end{figure}

Both the wavefunction and the correlator admit expansion in terms of Feynman graphs, see figure~\ref{figs:feyn}. We will focus most of our analysis at this graph level. For a graph $G$, with vertex parameters $x_v$ and edge parameters $y_e$, the wavefunction contribution takes the form (where we omit the loop integration):
\begin{align}\label{eq:wf_main}
\Psi_G = \int_{0}^{\infty} d t^{V} e^{-\sum_{v} x_v t_v} \prod_{e} G_{BB}(t_e,t'_e;y_e),
\end{align}
while the corresponding correlator integral is
\begin{align}\label{eq:corr_main}
\langle G \rangle = \int_{-\infty}^{\infty} d t^{V}\int  e^{-\sum_{v} x_v |t_v|} \prod_{e} G_F(t_{e},t'_{e}; y_e).
\end{align}
Here the bulk-to-bulk propagator decomposes as
\begin{align}
G_{BB}(t,t';y) = G_F(t_e,t'_e;y)-H(t,t';y),
\label{eq:GBB}
\end{align}
where we have
\begin{align}
G_F = \frac{1}{2 y}e^{-y|t-t'|} =\frac{1}{2 y} \widehat{G}_F, \quad H= \frac{1}{2 y} e^{-y(t+t')},
\end{align}
with $G_F$ the Feynman propagator and $H$ enforcing the vanishing condition of the bulk-to-bulk propagator when either endpoint is taken to the boundary. We emphasize that equation \eqref{eq:corr_main} is the textbook expression for a time-ordered correlation function. When all points are space-like separated, the operators all commute, and the time-ordering is irrelevant; thus, the expression yields the in-in correlator. 
At the graph level, there is growing evidence to support this philosophical hope for simplification. First, correlators exhibit only a subset of the singularities present in the wavefunction \cite{Glew:2025mry}, drastically reducing the complexity of the relevant function space \cite{Chowdhury:2023arc,Benincasa:2024ptf}.  Second, the `melonic' loop integrands behave much like their tree-level counterparts \cite{Donath:2024utn}. A key structural reason for these simplifications is that the wavefunction is computed over a half-time domain, from the asymptotic past up to the boundary. In contrast, the correlator arises from integral over the full time domain \footnote{The equivalence between in-in formalism and in-out correlation functions has been discussed in \cite{Kamenev2011} for flat space, and \cite{Donath:2024utn} for de Sitter correlators. Additionally, here we work in Euclidean signature, the equivalence between Lorentzian de Sitter in-in correlators and Euclidean de Sitter was discussed in \cite{Higuchi:2010xt}}. Here, we develop these observations and present explicit computations highlighting the relative simplicity of correlators. We will keep the discussion in flat space; however, many of these simplifications carry over to a de Sitter background. Additionally, we will restrict ourselves to theories with only polynomial interactions, also in this case we expect that the simplifications generalize for theories with derivative interactions. Our main results are:
\newline
\noindent{\bf Soft limits}---At tree level when taking the external energies to zero, except those in a connected sub-graph, the correlator factorizes into a product of the amplitude obtained by shrinking the sub-graph to a vertex and the correlator of the sub-graph. In addition, at any loop order, we find that taking all but one of the vertex energies to zero, the correlator reduces to the $\ell_0$-integrated amplitude \footnote{By $\ell_0$-integrated amplitude we are referring to the integrand where the spatial loop integration is still undone} up to a factor. Furthermore, when taking any two external energies to zero in a tree-level correlator, it's behaviour can be recovered from a simple contour integral. 
\newline
\noindent{\bf  Systematic expansion}---Although the factorisation properties of the wavefunction and correlator are closely related, we show that expansions around a given pole are systematically simpler for correlators. For the correlator, the first sub-leading term always vanishes. Furthermore, for the total-energy pole, the next two subleading terms take remarkably compact forms that can be generated by acting with a simple differential operator on the ($\ell_0$-integrated) amplitude.

\section{Wavefunction vs correlator}
We work at the level of individual diagrams. It is convenient to introduce variables tailored to the truncated graph $G$, obtained from the original Feynman graph by deleting all external edges. To each vertex $v \in V(G)$ we assign a variable $x_v$, defined as the sum of the magnitude of the external momenta entering the vertex. Likewise, to each edge $e \in E(G)$ we assign a variable $y_e$, corresponding to the magnitude of the momentum flowing through it. Consider the five-point tree-level process shown in Fig.~\ref{figs:feyn}, the associated graph variables are:
\begin{align}
x_1 &= |{\bf k}_1|+|{\bf k}_2|,  \quad x_2 = |{\bf k}_3|,  \quad x_3 = |{\bf k}_4|+|{\bf k}_5|, \notag \\
y &= |{\bf k}_1+{\bf k}_2|=|{\bf k}_3+{\bf k}_4+{\bf k}_5|, \notag \\
\tilde{y} &= |{\bf k}_4+{\bf k}_5|=|{\bf k}_1+{\bf k}_2+{\bf k}_3|.
\end{align}
We will also rescale the wavefunction and correlator as 
\begin{align}
\Psi(G) = \frac{1}{\mathcal{N}_G} \widehat{\Psi}(G), \quad \langle G \rangle = \frac{1}{\mathcal{N}_G} \langle \widehat{G} \rangle,
\end{align}
where $\mathcal{N}_G = \prod_{e \in E(G)} 2y_e$.
\newline
\newline
\noindent{\bf Wavefunction singularities}---The wavefunction develops singularities whenever the total energy entering \emph{any} connected subgraph $\mathfrak{g} \subseteq G$ vanishes \cite{CosmoPolys}. In this context, we use the convention that any edge entering the sub-graph from outside, is replaced by a half-edge attached to the corresponding vertex, with the half-edge inheriting the edge variable of the original. For instance, consider the sub-graph of the two-cycle shown in figure~\ref{figs:feyn} consisting of only the top edge, for which we have: 
\begin{align}\label{eq:graph_conv}
\mathfrak{g} = \raisebox{-0.33cm}{\includegraphics[scale=0.9]{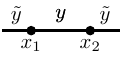}}.
\end{align}
In the graph variables $\{ x_v,y_e\}$, the singularity associated to each connected subgraph $\mathfrak{g} \subseteq G$ is:
\begin{align}
\mathcal{E}(\mathfrak{g}) = \sum_{v \in V(\mathfrak{g})} x_v + \sum_{e \in H(\mathfrak{g})}  y_e,
\label{eq:tube_func}
\end{align}
where $H(\mathfrak{g})$ is the set of half-edges of the subgraph. For the sub-graph considered in \eqref{eq:graph_conv}, the corresponding singularity is located at:
\begin{align}
\raisebox{-0.15cm}{\includegraphics[scale=0.6]{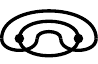}} =x_1+x_2+2 \tilde{y}.
\end{align}
Here we have depicted the subgraph by encircling it and introduced the shorthand $\raisebox{-0.15cm}{\includegraphics[scale=0.5]{figs/twocycle_tube_12.pdf}} \equiv \mathcal{E}(\raisebox{-0.15cm}{\includegraphics[scale=0.5]{figs/twocycle_tube_12.pdf}})$.
\newline
\newline
\noindent{\bf Wavefunction residues}---Given a connected subgraph $\mathfrak{g}$ let $\{ \mathfrak{h}_i\}$ be the connected components of the graph $G \setminus \mathfrak{g}$. For each component we define $\mathcal{Y}_i=H(\mathfrak{h}_i)$ as the set of edges with one endpoint in $\mathfrak{g}$ and the other in $\mathfrak{h}_i$. The behaviour of the wavefunction at the pole $\mathcal{E}(\mathfrak{g})$ is as follows:
\begin{align}
&\text{Res}_{\mathcal{E}(\mathfrak{g})=0} \widehat\Psi(G) = \mathcal{A}(\mathfrak{g})   \prod_{i} \sum_{\sigma \in \Sigma^{|\mathcal{Y}_i|}}(-1)^{\sigma_+} \widehat\Psi^\sigma(\mathfrak{h}_{i}),
\label{eq:WF_fact}
\end{align}
where $\Sigma^n = \{ +,-\}^n$ denotes the set of sign patterns of cardinality $n$. The shifted wavefunctions are defined as $\widehat\Psi^{\sigma}(\mathfrak{g})\equiv\widehat\Psi(\overline{\mathfrak{g}}_{i}, \mathcal{Y}_i \rightarrow \sigma \mathcal{Y}_i)$ where the replacement rule instructs us to flip the signs of the variables $y_e$ for $e \in \mathcal{Y}_i$ according to the sign pattern $\sigma$, and $\sigma_+$ denotes the number of plus signs in $\sigma$.
The factor $\mathcal{A}(\mathfrak{g})$ is the amplitude of the subgraph $\mathfrak{g}$, when we are considering loop sub-graphs then it is the integrand obtained after integrating the zeroth component of the loop momentum ($\ell_0$), although we refer to it as $\ell_0$-integrated amplitude.
\newline
\newline
\noindent{\bf Correlator singularities}---The correlator contains only the singularities associated with connected, vertex-induced subgraphs of $G$, resulting in a pole structure that is considerably simpler than that of the wavefunction. This cancellation of poles mirrors the fact that the integrated wavefunction appears to have a more complicated function space than the correlator \cite{Chowdhury:2023arc, Benincasa:2024ptf}. For  example, the bubble singularities are: 
\begin{align}
\overbrace{\underbrace{\raisebox{-0.cm}{\includegraphics[scale=0.7]{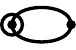}} \quad \raisebox{-0.cm}{\includegraphics[scale=0.7]{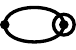}} \quad
\raisebox{-0.0cm}{\includegraphics[scale=0.7]{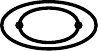}}}_{\text{correlator}} \quad
\raisebox{-0.cm}{\includegraphics[scale=0.7]{figs/twocycle_tube_12}} \quad \raisebox{-0.13cm}{\includegraphics[scale=0.7]{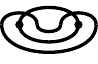}}}^{\text{wavefunction}},
\end{align}
That only poles of the type $\sum_v x_v +\sum_e y_e$ appear, and never with $2 y_e$, is true for any correlator, and is seen directly from the integral representation \eqref{eq:corr_main}.  All we need is to isolate one edge integration, pick one of the terms in $G_F$ (\textit{e.g.:} $\theta(t_i-t_j)$), and isolate the integration regions where the vertex times have opposite sign. One part is: 
\begin{equation}
    \int_{-\infty}^{\infty} dt_i dt_j e^{x_i t_i} e^{- x_j t_j} e^{-y(t_i - t_j)} \theta(t_i-t_j) \theta(-t_i) \theta(t_j) = 0 \, ,
\end{equation}
The other part is:
\begin{equation}
    \int_{-\infty}^{\infty} dt_i dt_j e^{-x_i t_i} e^{x_j t_j} e^{-y(t_i - t_j)} \theta(t_i-t_j) \theta(t_i) \theta(-t_j)\, ,
\end{equation}
which once we enforce the condition from $\theta(t_i)$ and $\theta(-t_j)$, the step function $\theta(t_i-t_j)$ becomes automatically satisfied, and no $2y$ can appear in the exponential.
\newline
\newline
\noindent{\bf Correlator residues}---At tree level, the correlator exhibits a factorisation property analogous to that of the wavefunction: taking the residue at the pole $\mathcal{E}({\mathfrak{g}})$ yields 
\begin{align}
&\text{Res}_{\mathcal{E}(\mathfrak{g})=0}\langle \widehat G\rangle = \mathcal{A}(\mathfrak{g})\times\prod_{i}  \sum_{\sigma \in \Sigma^{|\mathcal{Y}_i|}}  \langle \hat{\mathfrak{h}}_i\rangle^\sigma,
\end{align}
where the shifted correlators are defined as for the wavefunction $ \langle \hat{\mathfrak{h}}_i\rangle^\sigma \equiv \langle \hat{\mathfrak{h}}_i, \mathcal{Y}_i \rightarrow \sigma \mathcal{Y}_i\rangle$. At loop level, however, the residue structure is no longer straightforward, making the simplified expansion for the correlator presented in Section~\ref{sec:expand} all the more remarkable.
\newline
\newline
\noindent{\bf Melonic simplifications}---As pointed out in \cite{Donath:2024utn}, for a graph with multiple parallel edges, the correlator can expressed in terms of a single effective edge as
\begin{align}
\left \langle \raisebox{-0.47cm}{\includegraphics[scale=0.8]{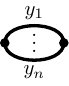}} \right \rangle  =\frac{2 y_T} {\prod_{i=1}^n (2y_i)} \left \langle \raisebox{+0.03cm}{\includegraphics[scale=0.8]{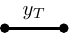}} \right \rangle, \text{with}\:\:  y_T=\sum_{i=1}^n y_i .
\end{align}
This allows us to restrict our analysis of the correlator to graphs in which each pair of vertices is connected by at most one edge.
\section{Soft Limits}

\noindent{\bf Time-integrals for rooted trees}---At tree-level, we may choose an arbitrary root vertex $v_*$ such that, for each $v \in V(G)$, there exists a unique path $P_*^v$ connecting $v_*$ to $v$. It is then convenient to introduce the edge variables
\begin{align}
s_e = t_e^{\uparrow}-t_e^{\downarrow},
\label{eq:var_chg}
\end{align}
where $t_e^{\uparrow}$ denotes the time variable associated with the endpoint of $e$ further from $v_*$, and $t_e^{\downarrow}$ the time variable associated to the endpoint of $e$ closer to $v_*$. The translation between the edge variables $s_e$ and the time variables appearing in \eqref{eq:corr_main} is given by 
$
t_v = t_* + S_v \quad \text{ with } \quad S_v = \sum_{e \in P_*^v} s_e.
$
Making these substitutions, \eqref{eq:corr_main} becomes
\begin{align}\label{eq:corr_root}
\langle \widehat G \rangle = \int_{-\infty}^{\infty} d t_*\int_{-\infty}^{\infty} d s^{E} e^{-\sum_{v} x_v |t_*+S_v|}\prod_{e \in E(G)} \widehat{G}_F(s_e).
\end{align}

\noindent{\bf Connected subgraphs}---We now explore the case where we take all vertex variables in a graph $G$ to zero, except those in a connected subgraph $\mathfrak{g}$ \footnote{Here we are treating site variables, $x_v$, to be independent of the edge variables, $y_e$, so we can formally take the limit.}. Choosing $x_{v_*}$, where $v_* \in \mathfrak{g}$, as the root in \eqref{eq:corr_root}, then the integral factorizes into the correlator of $\mathfrak{g}$, and the amplitude of a graph where $\mathfrak{g}$ is contracted to a single vertex, which we denote by $G/\mathfrak{g}$.: 
\begin{align}
\langle \widehat G \rangle &= \langle \hat{\mathfrak{g}}  \rangle \times  \tilde{\mathcal{A}}({G/\mathfrak{g}}) ,\;  \tilde{\mathcal{A}}({G/\mathfrak{g}}) = \int\limits_{-\infty}^{\infty} d s_{e} \prod_{e \in E(G/\mathfrak{g})} \widehat{G}_F(s_e) \nonumber\\
\langle \hat{\mathfrak{g}}  \rangle &= \int\limits_{-\infty}^{\infty}d t_* d s_{e^{\prime}} e^{-\sum_{v\in \mathfrak{g}} x_v |t_*+S_v|} \prod_{e^{\prime} \in E(\mathfrak{g})} \widehat{G}_F(s_{e^{\prime}})]\,.
\label{eq:softlim}
\end{align}
Where $\tilde{\mathcal{A}}(G)$ is the off-shell, amputated, time-ordered correlator, evaluated at zero external energies, which we refer to as amplitude for simplicity. When $\mathfrak{g}$ is a single vertex, then the property holds for loop graphs too, and we obtain the $\ell_0$-integrated amplitude:
\begin{align}
\langle \widehat G \rangle |_{x_{v_*} \neq 0}  = \frac{2}{x_{v_*}} \times \tilde{\mathcal{A}}(G),
\end{align}
For a connected graph, the \(l_0\)-integrated amplitude $\mathcal{\tilde{A}}(G)$ satisfies a Berends-Giele type recursion
\begin{align}
\tilde{\mathcal{A}}(G)
= \sum_{v \in V(G)} \prod_{\mathfrak{g} \in \kappa(G \setminus v)}
\left.
\frac{\tilde{\mathcal{A}}(\mathfrak{g})}{\mathcal{E}(\mathfrak{g})}
\right|_{x_v = 0},
\end{align}
where the product runs over the connected components of the graph \(G \setminus v\) and the $\mathcal{E}(G)$ are the usual tube functions defined in \eqref{eq:tube_func} evaluated at $x_v=0$.
The recursion terminates on graphs consisting of a single vertex, for which we have \( \tilde{\mathcal{A}}(v)=1\).

As an example, consider the three-site chain. Applying the recursion once gives
\begin{align}
\tilde{\mathcal{A}}  \left( \vcenter{\hbox{\includegraphics[scale=0.7]{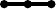}}} \right) &= \frac{\tilde{\mathcal{A}}  \left( \vcenter{\hbox{\includegraphics[scale=0.7]{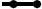}}} \right)}{\mathcal{E}  \left( \vcenter{\hbox{\includegraphics[scale=0.7]{figs/path_amp_rec_2.pdf}}} \right)}+\frac{1 }{\mathcal{E}  \left( \vcenter{\hbox{\includegraphics[scale=0.7]{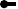}}} \right)} \frac{1}{\mathcal{E}  \left( \vcenter{\hbox{\includegraphics[scale=0.7]{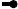}}} \right)}+\frac{\tilde{\mathcal{A}}  \left( \vcenter{\hbox{\includegraphics[scale=0.7]{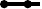}}} \right)}{\mathcal{E}  \left( \vcenter{\hbox{\includegraphics[scale=0.7]{figs/path_amp_rec_3.pdf}}} \right)},
\end{align}
where the two-vertex paths themselves recurse as
\begin{align}
\tilde{\mathcal{A}}  \left( \vcenter{\hbox{\includegraphics[scale=0.7]{figs/path_amp_rec_2.pdf}}} \right) = \frac{1}{\mathcal{E}  \left( \vcenter{\hbox{\includegraphics[scale=0.7]{figs/path_amp_rec_2_r.pdf}}} \right)}+\frac{1}{\mathcal{E}  \left( \vcenter{\hbox{\includegraphics[scale=0.7]{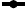}}} \right)}, \notag \\
\tilde{\mathcal{A}}  \left( \vcenter{\hbox{\includegraphics[scale=0.7]{figs/path_amp_rec_3.pdf}}} \right) = \frac{1}{\mathcal{E}  \left( \vcenter{\hbox{\includegraphics[scale=0.7]{figs/path_amp_rec_2_lr.pdf}}} \right)}+\frac{1}{\mathcal{E}  \left( \vcenter{\hbox{\includegraphics[scale=0.7]{figs/path_amp_rec_2_l.pdf}}} \right)}.
\end{align}
Using the definition of $\mathcal{E}(\mathfrak{g})$ in \eqref{eq:tube_func} for each connected subgraph, this gives
\begin{align}
\tilde{\mathcal{A}} = \frac{\frac{1}{\tilde y}+\frac{1}{y + \tilde y}}{y}+\frac{1}{y}\frac{1}{\tilde y}+\frac{\frac{1}{y+ \tilde y}+\frac{1}{y}}{\tilde y} = \frac{4}{y \tilde{y}}.
\end{align}
Notice that the recursion introduces the spurious pole $y+\tilde y=0$ which cancels in the sum. 
\newline\newline
\noindent{\bf Pairwise soft limits}---Now we consider the limit where we take all but two vertex variables ($x_*$ and $x$) to zero, in a tree graph $G$. In this limit the integral becomes
\begin{align}
&\langle \widehat G \rangle = \left(\int\limits_{-\infty}^{\infty} d s_{e^{\prime}} \prod_{e^{\prime} \in E(G/P_*)} \widehat{G}_F(s_{e^{\prime}})\right)\times \nonumber \\
&\times \left(\int\limits_{-\infty}^{\infty} d t_* d s_{e} e^{-x_* |t_*|-x |t_* + S_*|}\prod_{e \in E(P_*)} \widehat{G}_F(s_e)\right) \, .
\label{eq:2ptfunc_1}
\end{align}
The first integral is simply an amplitude and we are going to get a product over $\frac{1}{y}$. The second integral can be computed by a very simple contour integral, and the result is in fact very simple (check the appendix \ref{app:derivations} for the derivation), we obtain the full result as:

\begin{align}
&\langle \widehat G \rangle = \left(\prod_{e \in E(G/P_*)} \frac{2}{y_e}\right) \times \nonumber \\
&\sum_{p_* \in \{x_*,x, y_e\}}2 i\mathop{\mathrm{Res}}_{p\to i p_*}\frac{2 x_* x \prod_{e \in E(P_*)}2 y_e}{(p^2+x_*^2)(p^2+x^2)\prod_{e \in E(P_*)}(p^2+y_e^2)}.
\end{align}
\section{Expanding the correlator}
\label{sec:expand}
\noindent{\bf Tree-level $\mathcal{O}(\mathcal{E}_G)$-expansion}---We begin by examining the expansion around $\epsilon =0$, when we rescale the variables $x_v = \epsilon \bar{x}_v$ and $\tau = \epsilon t_*$, such that we are effectively expanding around the total energy pole. Then, \eqref{eq:corr_root} becomes
\begin{align}
\langle \widehat{G}  \rangle = \frac{1}{\epsilon}\int_{-\infty}^{\infty} d \tau\int_{-\infty}^{\infty} d s^{E} F(\epsilon) \prod_{e \in E(G)} \widehat{G} _F(s_e),
\label{eq:Corr_varchg}
\end{align}
where we have defined
\begin{align}
F(\epsilon)= e^{-\sum_{v} \bar{x}_v |\tau+ \epsilon S_v|}.
\end{align}
The expansion of the correlator around $\epsilon=0$ is governed by the derivatives of $F(\epsilon)$, with the first few given explicitly by
\begin{align}
F'(0) &= -F(0)\sum_{v} \bar{x}_v S_v \text{sgn}(\tau) , \notag \\
F''(0) &= F(0) \Bigg\{ \left(\sum_{v} \bar{x}_v S_v \right)^2  -2 \sum_{v} \bar{x}_v S_v^2 \delta(\tau)  \Bigg \}.
\label{eq:Et1exp}
\end{align}
For instance, the leading order term is determined by the integral 
\begin{align}
\langle  \widehat{G}  \rangle^{(-1)} = \int_{-\infty}^{\infty} d \tau F(0) \int_{-\infty}^{\infty} d s^{E}  \prod_{e \in E(G)} \widehat{G}_F(s_e).
\end{align}
In analogy to the discussion around \eqref{eq:corr_root}, the above integrals can be performed  to find
\begin{align}
\langle  \widehat{G}  \rangle^{(-1)}  =\frac{2}{\sum_{v} \bar{x}_v}  \prod_{e \in E(G)} \frac{2}{y_e}.
\label{eq:Ord1_Op}
\end{align}
The correlator at order $\mathcal{O}(\epsilon^0)$ is trivial to evaluate. Since $F'(0)$ contains a factor of $\text{sgn}(\tau)$ the integrand becomes odd in $\tau$, and the integral consequently vanishes, leading to 
\begin{align}
\langle \widehat G \rangle^{(0)} = 0. 
\end{align}
This vanishing at next-to-leading order is unique to the correlator, making it significantly simpler than the wavefunction! The relative simplification continues to the next term in the expansion:
\begin{align}
\langle \widehat{G} \rangle^{(1)} = \frac{-1}{2\sum_{v} \bar{x}_v}\sum_{u \neq v} \bar{x}_u \bar{x}_v \int_{-\infty}^{\infty} d s^{E}  \sum_{e \in P_u^v} s_e^2   \prod_{vv' \in E(G)} \widehat{G}_F(s_e),
\label{eq:tree_Et1term}
\end{align}
where $P_u^v$ is the unique path from vertex $u$ to $v$, see appendix \ref{app:derivations}. The \( s_e^2 \) factor is generated by the action of a suitable differential operator on the amplitude, leading to the final form of the \( \mathcal{O}(\epsilon) \) contribution to the expansion:
\begin{align}
\langle \widehat G \rangle^{(1)} =- \sum_{u \neq v} \frac{\bar{x}_u \bar{x}_v}{\mathcal{E}_G} \sum_{e \in P_u^v} \partial_{y_e}^2  \prod_{e \in E(G)} \frac{2}{y_e},
\label{eq:tree_Et1termOP}
\end{align}
where the total energy is given by $\mathcal{E}_G = \sum_{v}\bar{x}_v$.
\newline\newline
\noindent{\bf Loop Correlators $\mathcal{O}(\mathcal{E}_G)$-expansion}---At loop-level, our change of variables \eqref{eq:var_chg} imposes relations between the $s_e$ variables. For example, for one-loop $n$-gon diagrams, under the support of \eqref{eq:var_chg}, for a single variable $s_{e^{\prime}}$ we must impose the constraint:
\begin{equation}
s_{e^{\prime}}=-\sum_{e \in \mathcal{E}/e^{\prime}} s_e.
\label{eq:se_constraint}
\end{equation}
For an $n$-loop diagram $n$ variables $s_e$ will be constrained in terms of the others. The equations \eqref{eq:Et1exp} still hold true, and we recover the $\ell_0$-integrated amplitude in this limit for the $\mathcal{E}_G^{-1}$ coefficient, and the $\mathcal{E}_G^{0}$ coefficient also vanishes. However, for the $\mathcal{E}_G$ there is a slight complication. We find from \eqref{eq:path_expEt1}, that the cross terms of the form $s_{e} s_{e'}$ do not vanish, given that the integral is no longer odd. This is simply because the argument of the Feynman propagators will be \eqref{eq:se_constraint}. Therefore there is no obvious form for the differential operator acting on the amplitude which gives us the $\mathcal{E}_G$ term. Nevertheless, it is still possible to generate this coefficient. Because of \eqref{eq:se_constraint}, we can choose how we parametrize the sum of $s_e$ in \eqref{eq:path_expEt1}. This is because for loop diagrams, there is more than one path between two sites. Then, we can consider a one higher-loop graph where we added an edge between two sites, for this diagram  we can parametrize \eqref{eq:path_expEt1} such that the path between these two sites is a single $s_e$, and we do not have the issue of cross terms not vanishing. Then, for loop graphs the differential operator is simply:
\begin{equation}
    \langle \widehat G \rangle^{(1)} = - \frac{1}{2}\frac{1}{\sum_{v} \bar{x}_v}\sum_{u \neq v} \bar{x}_u \bar{x}_v  \partial_{y_{uv}}^2 \tilde{\mathcal{A}}^{uv}(G)\vert_{y_{uv}\to 0},
\end{equation}
where $\mathcal{A}^{uv}(G)$ is the $\ell_0$-integrated amplitude of the diagram with at least one edge connecting the sites $x_u$ and $x_v$. This is obvious from \eqref{eq:Corr_varchg}, adding one edge corresponds to adding one Feynman propagator, when we set the additional edge to zero after acting with the differential operator, we recover \eqref{eq:path_expEt1}.
\newline
\newline
\noindent{\bf $\mathcal{O}(\mathcal{E}_G^2)$-expansion}---The $\mathcal{O}(\mathcal{E}_G^2)$ term still exhibits a much simpler structure in the correlator:
\begin{align}
\langle \widehat G \rangle^{(2)} = \int_{-\infty}^{\infty} d s^{E} \sum_{u \neq v} \bar{x}_u \bar{x}_v  \left| \sum_{e \in P_u^v} s_e \right|^3 \sum \prod_{e \in E(G)} \widehat{G}_F(s_e)\, .
\end{align}
Due to the absolute value operator in the sum of $s_e$, even for tree level the cross terms no longer vanish. However, we can still apply the solution we used for loops, we add an edge between any two sites and then we can lower the powers of $s_e$ by the action of a single differential operator. The $\mathcal{O}(\mathcal{E}_G^2)$ term becomes the following: 
\begin{equation}
    \langle \widehat G \rangle^{(2)} = - \frac{1}{\sum_{v} \bar{x}_v}\sum_{u \neq v} \bar{x}_u \bar{x}_v  \partial_{y_{uv}}^3 \tilde{\mathcal{A}}^{uv}(G)\vert_{y_{uv}\to 0} .
\end{equation}
Beyond $\mathcal{O}(\mathcal{E}_G^2)$ the coefficients become complicated polynomials of $\bar{x}_v$, and it is not clear whether we can obtain them by acting with a differential operator on an amplitude. 
\newline

\noindent{\bf Partial energy pole expansion}---The correlator also admits a simple expansion around the partial energy poles. For a fixed connected subgraph $\mathfrak{g}$, we rescale all variables that appear in the pole as follows:
\begin{align}
& x_v \rightarrow \epsilon \bar{x}_v \text{ for all } v \in V(\mathfrak{g}), \notag \\
&y_e \rightarrow \epsilon \bar{y}_e \text{ for all } e \in H(\mathfrak{g}),
\end{align}
such that the partial energy becomes $\mathcal{E}_\mathfrak{g} \rightarrow \epsilon \mathcal{E}_{\mathfrak{g}}$. We can then derive the expansion by performing the change of variables described in \eqref{eq:var_chg} solely for the edges $e \in E(\mathfrak{g})$, and one time variable. At leading order we find the usual factorisation
\begin{align}
\langle \widehat{G} \rangle^{(-1)} = \frac{\tilde{\mathcal{A}}(\mathfrak{g}) }{\mathcal{E}_\mathfrak{g}}\prod_{\mathfrak{h} \in \kappa(G \setminus \mathfrak{g})}\langle \mathfrak{h} \rangle,
\end{align}
where the product is over all connected components of the graph $G$ with $\mathfrak{g}$ removed. In this limit, $\langle \mathfrak{h} \rangle$ is the correlator of the respective connected component  with the edges connected to $\mathfrak{g}$ set to zero. Furthermore, the first derivative with respect to $\epsilon$ is the same as in \eqref{eq:Et1exp}, therefore the sub-leading term in the series expansion around any pole of the correlator vanishes. This places strongly constraints the correlator. We found in tree-level examples up to $4$ vertices ($4$-chain
and star) that if we make an ansatz for the correlator inputing only the poles, demanding only the vanishing of subleading terms fixed $97\%$ of the coefficients.

\section{$\text{Tr}(\phi^3)$ correlator}

\noindent{\bf Soft Limits}---Recall that for a single graph, setting all vertex variables to zero except one, the correlator reduces to the corresponding contribution to the amplitude up to an overall factor. To extend this to the full correlator, we recall that the vertex variables are expressed as sums over subsets of consecutive energies. Consequently, for each graph, exists a unique vertex variable $x_{v*}$ whose sum includes the energy $|k_*|$. If we now switch off all other energies, this sets all vertex variables to zero except $x_{v_*}$ across all graphs simultaneously. The resulting expression is therefore
 \begin{align}
\langle \widehat{\phi}^n \rangle|_{|{\bf k}_{*}| \neq 0} = \frac{2}{|{\bf k}_*|} \tilde{\mathcal{A}}_n,
 \end{align}
where $\tilde{\mathcal{A}}_n$ is the $n$-point tree-level $\tr(\phi^3)$ amplitude (in the same sense as \eqref{eq:softlim}).
\newline\newline
\noindent{\bf Pole expansion}---The interesting structure in the expansions above generalizes for the sum of graphs. To obtain the expansion around a given pole we start by re-scaling every momenta contributing to the pole, $\vec{k}_i \to \epsilon \, \vec{k}_i$, and then expand around $\epsilon = 0$. The leading term will be the usual factorization structure of the correlator, and the sub-leading term will vanish. For the total energy pole the $\mathcal{O}(\mathcal{E}_{\text{tot}}^1)$ term will also have an interesting structure. We can reformulate our rule to obtain the operator acting on the amplitude to retrieve the $\mathcal{E}_{\text{tot}}$ term by thinking about each graph in terms of the momentum polygon for an $n$-point process. We know each triangulation of the polygon is dual to a graph, and the internal chords (parametrized by $|k_{ij}|$ of a triangulation are dual to edges (above parametrized by $y_i$) of the graph. Therefore, the rule of the $\mathcal{E}_{\text{tot}}$ operator is now that for each monomial $|\vec{k}_i|\,|\vec{k}_j|$ we take the sum over all the double derivatives with respect to the internal chords crossed when we draw a straight line from the outer edge $|\vec{k}_i|$ to the outer edge $|\vec{k}_j|$. This allows us to re-write \eqref{eq:Ord1_Op} as:
\begin{equation}
\nabla_{i}^j = -|\vec{k}_i|\,|\vec{k}_j| \sum_{e \in P_i^j} \partial_{|k|_e}^2,
\label{eq:Ord1_Op_full}
\end{equation}
where now $P_i^j$ is the set of chords which are crossed when we draw a straight line from $|\vec{k}_i|$ to $|\vec{k}_j|$ in the momentum polygon. This operator now acts on the amplitude of the full process $A_n$.
\section{Outlook}
In this letter, we explained how the relative simplicity of correlators follows from the basic fact that they are full spacetime integrals, while wavefunctions are defined only on half of spacetime. Thus the most physical observable---the correlator, obtained from the wavefunction via the Born rule---is both `simpler' and more directly tied to flat-space scattering amplitudes than the wavefunction. This picture exposes remarkable properties obscured in the in-in formalism, and opens some obvious directions. One striking feature is the vanishing of subleading terms in the Laurent expansion around {\it every} pole. This gives strong constraints for directly bootstrapping the correlator, whose power we plan to analyse systematically. Two natural generalizations are to correlators in de Sitter, and to derivative interactions and spinning correlators. Since the origin of simplicity--full spacetime integration--remains, it will be interesting to understand the cosmological avatar of these pole expansions. Finally, the improved behaviour near the $\mathcal{E}_{\text{tot}}$ pole has an interesting implication for Tr $\phi^3$ theory. If, instead of sending the total energy to zero, we shift all internal chords of the momentum polygon to infinity, the leading behaviour gives the amplitude, with vanishing subleading correction. For Tr $\phi^3$ amplitudes, hidden ``projective" symmetries under special $g$-vector shifts kill the leading behaviour, while the subleading term reveals a ``world at infinity" of other amplitudes \cite{ABHY,Zeros,Arkani-Hamed:2024nhp,HiddenZerosWF}. Starting from the correlator, however, {\it both} terms vanish, making the analogue of this ``world at infinity" fascinating.
\newline

\noindent{\bf Acknowledgements}---We thank Carolina Figueiredo, Daniel Baumann and Guilherme Pimentel for useful discussions. The work of N.A.H. is supported by the DOE (Grant No. DE-SC0009988), the Simons Collaboration on Celestial Holography, the ERC UNIVERSE+ synergy grant, and the Carl B. Feinberg cross-disciplinary program in innovation at the IAS. The work of F.V. is supported by the Jonathan M. Nelson Center for Collaborative Research. 

\newpage
\onecolumngrid
\appendix

\section{Some Derivations}
\label{app:derivations}

\noindent{\bf Two-point Function contour integral}--- Here we compute the second integral in \eqref{eq:2ptfunc_1}, which is:
\begin{align}
\mathcal{I}= \int\limits_{-\infty}^{\infty} d t_* d s_{e} e^{-x_* |t_*|-x |t_* + S_*|}\prod_{e \in E(P_*)} \widehat{G}_F(s_e)
\end{align}
By considering the Fourier transform:
\begin{equation}
    e^{-x |t_* + S_*|} = \int\limits_{-\infty}^{\infty}\frac{dp}{\pi}\frac{x}{p^2+x^2}e^{i p (t_*+S_*)}\, .
\end{equation}
Then, every integral over $t_*$ and $s_e$ can be directly computed, for example:
\begin{equation}
    \int\limits_{-\infty}^{\infty} d s_e\, e^{-y|s_e| + i p s_e}=\frac{2 y}{p^2+y^2}\, ,
\end{equation}
therefore the first integral in \eqref{eq:2ptfunc_1} becomes:
\begin{equation}
    \mathcal{I}=\int\limits_{-\infty}^{\infty}\frac{dp}{\pi}\frac{2 x_* x \prod_{e \in E(P_*)}2 y_e}{(p^2+x_*^2)(p^2+x^2)\prod_{e \in E(P_*)}(p^2+y_e^2)} \, .
\end{equation}
This integral is very easy to compute given that all of its poles live on the imaginary axis, we can simply compute the integral over the contour which encloses every pole in the positive imaginary axis (or equivalently all the ones in the negative axis). The integral over the arc clearly vanishes at $\pm i \infty$. Picking to contour over the positive imaginary axis, the result is a sum over all the residues on $p=i x_*$, $p=i x$ and $p=i y_e$:
\begin{equation}
    \mathcal{I}=2 i \sum_{p_* \in \{x_*,x, y_e\}}\mathop{\mathrm{Res}}_{p\to i p_*}\frac{2 x_* x \prod_{e \in E(P_*)}2 y_e}{(p^2+x_*^2)(p^2+x^2)\prod_{e \in E(P_*)}(p^2+y_e^2)}
\end{equation}
\noindent{\bf $\mathcal{O}(E_t)$-expansion term derivation}---Starting from \eqref{eq:Corr_varchg} the sub-sub-leading term is, explicitly:

\begin{align}
\langle G \rangle^{(1)} = \frac{\epsilon}{2}\int_{-\infty}^{\infty} d \tau\int_{-\infty}^{\infty} d s^{E} F(0) \Bigg\{ \left(\sum_{v} \bar{x}_v S_v \right)^2  -2 \sum_{v} \bar{x}_v S_v^2 \delta(\tau)  \Bigg \} \prod_{e \in E(G)} G_F(s_e),
\end{align}
performing the integral over $\tau$ we find that the delta function forces the integral to be $1$, and in the first term we obtain a pole over the sum of all vertex variables:
\begin{align}
\langle G \rangle^{(1)} = \epsilon \frac{1}{\sum_v \bar{x}_v} \int_{-\infty}^{\infty} d s^{E} \left\{ \left(\sum_{v} \bar{x}_v S_v \right)^2  -\sum_{u} \bar{x}_u \sum_{v} \bar{x}_v S_v^2 \right\}  \prod_{e \in E(G)} G_F(s_e).
\label{eq:path_expEt1}
\end{align}
The expression in curly brackets becomes: 
\begin{equation}
\left(\sum_v \bar{x}_v S_v\right)^2
-
\sum_{u,v} \bar{x}_u \bar{x}_v S_v^2
=
-\frac{1}{2}
\sum_{u,v} \bar{x}_u \bar{x}_v
\left(S_u-S_v\right)^2
=
-\frac{1}{2}
\sum_{u\neq v} \bar{x}_u \bar{x}_v
\left(
\sum_{e\in P_u^v} s_e
\right)^2
\end{equation}
where we have introduced the notation $P_u^v$ for the path between $u$ to $v$. Thus, only edges on the path from $u$ to $v$ can enter the sum on the RHS. Considering that the cross terms vanish upon integration, we are finally left with 
\begin{align}
\langle \widehat G \rangle^{(1)} =-\frac{1}{2} \frac{\epsilon}{\sum_{v} \bar{x}_v}\sum_{u \neq v} \bar{x}_u \bar{x}_v \int_{-\infty}^{\infty} d s^{E}  \sum_{e \in P_u^v} s_e^2   \prod_{vv' \in E_G} \widehat{G}_F(s_e),
\end{align}

\bibliographystyle{apsrev4-1.bst}
\bibliography{Refs.bib}

\end{document}